\documentclass[a4paper,12pt]{article}
\pdfoutput=1
\usepackage{graphicx, rotating, slashed, color, colortbl}

\usepackage[usenames,dvipsnames]{xcolor}
\usepackage{graphicx}
\usepackage{colortbl}
\usepackage[unicode=true,pdfusetitle,
 bookmarks=true,bookmarksnumbered=false,bookmarksopen=false,citecolor=Turquoise,linktoc=page,
 breaklinks=false,pdfborder={0 0 1},backref=false,colorlinks=true,pdfpagemode=FullScreen]
 {hyperref}
\usepackage{amssymb}
\newcommand{\bea}{\begin{eqnarray}}
\newcommand{\eea}{\end{eqnarray}}
\newcommand{\be}{\begin{equation}}
\newcommand{\ee}{\end{equation}}
\newcommand{\bc}{\begin{center}}
\newcommand{\ec}{\end{center}}
\newcommand{\bit}{\begin{itemize}}
\newcommand{\eit}{\end{itemize}}

\newcommand{\met}{\textrm{ mET}}
\newcommand{\jets}{\textrm{ jets}}
\newcommand{\leptons}{\textrm{ leptons}}
\newcommand{\ev}{\textrm{ eV}}

\newcommand{\gev}{\textrm{ GeV}}

\newcommand{\tev}{\textrm{ TeV}}
\newcommand{\TeV}{\textrm{ TeV}}
\newcommand{\eq}[1]{eq.~(\ref{#1})}
\newcommand{\eqs}[1]{eqs.~(\ref{#1})}

\newcommand{\vev}[1]{\langle #1 \rangle}

\usepackage{amsmath}
\usepackage{amssymb}
\usepackage{color}
\definecolor{rosso}{cmyk}{0,1,1,0.4}
\definecolor{rossos}{cmyk}{0,1,1,0.55}
\definecolor{rossoc}{cmyk}{0,1,1,0.2}
\definecolor{blu}{cmyk}{1,1,0,0.3}
\definecolor{blus}{cmyk}{1,1,0,0.6}
\definecolor{bluc}{cmyk}{1,1,0,0.1}
\definecolor{verde}{cmyk}{0.92,0,0.59,0.25}
\definecolor{verdec}{cmyk}{0.92,0,0.59,0.15}
\definecolor{verdes}{cmyk}{0.92,0,0.59,0.4}
\definecolor{grigio}{cmyk}{0,0,0,0.1}
\definecolor{rosa}{cmyk}{0,0.1,0.1,0.02}
\definecolor{rosino}{cmyk}{0,0.05,0.05,0.02}
\definecolor{rosas}{cmyk}{0,0.3,0.25,0.05}
\definecolor{celeste}{cmyk}{0.1,0,0,0.02}
\definecolor{giallino}{cmyk}{0,0,0.1,0.02}
\definecolor{rosso}{cmyk}{0,1,1,0.4}
\definecolor{rossos}{cmyk}{0,1,1,0.55}
\definecolor{rossoc}{cmyk}{0,1,1,0.2}
\definecolor{blu}{cmyk}{1,1,0,0.3}
\definecolor{blus}{cmyk}{1,1,0,0.5}
\definecolor{bluc}{cmyk}{1,1,0,0.1}
\definecolor{blucc}{cmyk}{0.7,0.5,0,0}
\definecolor{viola}{cmyk}{0,1,0,0.6}
\definecolor{viola2}{cmyk}{0,1,0.2,0.6}
\definecolor{verde}{cmyk}{0.92,0,0.59,0.25}
\definecolor{verdec}{cmyk}{0.92,0,0.59,0.15}
\definecolor{verdes}{cmyk}{0.92,0,0.59,0.4}
\definecolor{verdino}{cmyk}{0.12,0,0.09,0.02}
\definecolor{giallo}{cmyk}{0,0,1,0}
\definecolor{gialloverde}{cmyk}{0.44,0,0.74,0}

\providecommand{\tabularnewline}{\\}
\usepackage{array}
\usepackage{booktabs}
\usepackage{multirow}
\usepackage{caption3} 
\DeclareCaptionOption{parskip}[]{} 
\usepackage[small]{caption}
\newcommand{\yeff}{Y_{\textrm{eff}}}
\begin{document}
\begin{flushright}
UMD-PP-013-007\\
June 2013
\end{flushright}

\bigskip
\begin{center}
{\huge Radiatively Induced Type II seesaw and Vector-like 5/3 Charge Quarks\\} 
\vspace*{1cm}
{\bf \large R. Franceschini and R. N. Mohapatra\\}
\vspace{0.3cm}
{  \it Maryland Center for Fundamental Physics and Department of Physics\\ University of Maryland, College Park, MD 20742, USA}

\bigskip
~
\bigskip

\centerline{\large\bf Abstract}
\begin{quote}
\noindent Understanding small neutrino masses in type II seesaw models  with TeV scale SM triplet Higgs bosons  requires that its coupling with the standard model Higgs doublet $H$ be ``dialed'' down to be order eV to KeV, which is a fine-tuning by a factor of $10^{-11}-10^{-8}$ with respect to the weak scale. We present a SUSY extension of the type II seesaw model where this dimensionful small coupling is radiatively induced, thus making its smallness natural. This model has an exotic vector-like quark doublet  which contains a quark $X$ with electric charge $5/3$ and a top partner $t'$. 
We discuss in details the phenomenology of the model paying special attention to the consequences of the interactions of the the exotic heavy quarks and the scalars of the model. Implications for neutrinoless double beta decay and for the LHC experiments are discussed in detail. Remarkably, in this model both the seesaw triplet and the heavy quarks can manifest at colliders in a host of different signatures, including some that significantly differ from those of the minimal models. 
 Depending on the choice of the hierarchy of couplings, the decay of the heavy quarks and of the seesaw triplet may be subject to bounds that can be tighter or looser than the bounds from standard LHC searches.
Furthermore we point out  a new short-distance contribution to neutrinoless double beta decay mediated by the simultaneous propagation of the type II triplet and exotic fermions. {Remarkably this contribution to the  neutrinoless double beta decay is parametrically quite independent from the scale of the generated neutrino mass.} 
\end{quote}

\end{center}

\newpage

\tableofcontents

\bigskip

\section{Introduction}

Seesaw mechanism provides a simple way to understand the smallness of the neutrino mass in gauge theories. It relates the neutrino mass to the inverse of a high mass for new fields added to the standard model e.g.  gauge singlet fermion (in type I)~\cite{seesaw} or SM triplet Higgs boson in type II seesaw~\cite{seesaw2} case.  Their mass terms being SM singlets are unconstrained by the standard model (SM) gauge symmetry and are allowed to be as high as needed. For order one couplings of leptons to these new fields, this mass scale has to be in the range of $10^{14}$ GeV to get neutrino masses in the eV to sub-eV range. As a result, the presence of new fields and hence these simple seesaw mechanisms becomes hard to test  by any low energy or collider experiments except that with additional assumptions like supersymmetry, they may lead to indirect observable signals in processes such as lepton flavor violation, depending on the seesaw scale. With LHC searches for TeV scale new particles in full swing, the possibility that the new fermions and bosons connected with seesaw type I or II  may have masses in the TeV range has attracted considerable interest~\cite{typeii}. However, in such a situation, some parameters in the seesaw relation have to be dialed down to rather small values and tend to look less natural. It is therefore important to search for theoretical settings where the input parameters to the seesaw formula arises only as higher order effects so that a TeV scale would look natural and provide a stronger case for searches of the seesaw particles at LHC. In this article we present a model where a key lepton number violating parameter that goes into the type II seesaw formula for neutrino masses is naturally vanishing at the tree level and arises only at the one loop level so that it allows the associated seesaw bosons to have masses in the TeV range in a natural manner~\cite{earlier}.  The literature on TeV scale models where neutrino mass arises at the one or two loop levels, {outside the seesaw framework}, is enormous~\cite{loop} and there have also been extensive studies of higher dimensional operators and how they contribute to TeV scale models for eV scale neutrino masses~\cite{tev}.

Our model has several testable phenomenological consequences. A distinctive feature of the model with respect to ordinary seesaw models is the introduction of a new quark with electric charge 5/3 (denoted by $X_{5/3}$) as the weak isospin partner of a top partner $t'$. The $(X_{5/3},t')$ form a vector-like SM doublet, which has a gauge-invariant mass term. It is quite natural in the model for the $X_{5/3}$ particle to have TeV to sub-TeV masses so that it can be searched at the CERN LHC. 

The $t'$  has electric charge 2/3 and after symmetry breaking it mixes with the $u, c, t$ quarks of the SM. As it happens in standard scenarios of heavy quarks this mixing induces a  decay of $t'$ into SM weak gauge bosons and SM quarks. Through this same mixing, the $t'$ changes the couplings of the SM quarks to the weak vectors bosons. Both aspects of the $t'$ phenomenology have been extensively studied~\cite{q5/3}.

The $X_{5/3}$ particle has been discussed in other contexts~\cite{q5/3} such as little Higgs or composite Higgs models. Interestingly, here we are led to the same kind of particles from neutrino mass point of view. While the dominant decay mode of the $X_{5/3}$ particle can still be $$X_{5/3}\to tW,$$ as in most models where this particle is introduced, our type II seesaw model endows it with a significant branching ratio to other states of the Higgs sector, such as  $$X_{5/3}\to \Delta^{++} d_{i}\,,$$ which gives rise to multi-leptons final states such as $b\bar{b}\ell^+\ell^+\ell^-\ell^-\,,$ and 
$$X_{5/3}\to H^{+}_{d} u_{i}\,,$$ which gives rise to $tbu_{i}$ final states. We remark that these final states are quite different from the final states usually considered for the $X_{5/3}$  and $t'$ searches~\cite{Chatrchyan:2012af,ATLAS:2012qe,Chatrchyan:2012vu,q5/3cms,ATLAS-CONF-2013-018}. Therefore the heavy quarks of this model are an interesting and motivated incarnation of top partners with non-minimal phenomenology.

Remarkably, we also find that in our model there is a new type of contribution to $\beta\beta_{0\nu}$ decay which not only leads to constraints on the model parameters but also provides a new way to probe our idea.

\bigskip

The paper is organized as follows: in Sec.~\ref{model}, we present the extension of the supersymmetric type II seesaw model by adding vector-like quark doublets with $Y=\frac{7}{6}$ and show how this makes TeV scale type II natural; in Sec.~\ref{lhc}, we discuss phenomenological implications of the model at LHC and in Sec.~\ref{sec:doublebeta} we point out a new contribution to neutrinoless double beta decay that arises in this model. In Sec.~\ref{other} we briefly discuss precision flavor and electroweak observables. In Sec.~\ref{conclusions} we give our conclusions.

\section{5/3 charge quark and ``natural'' type II seesaw\label{model}} 

\subsection{Fine tuning problem in type II seesaw} Type II seesaw~\cite{seesaw2} mechanism for small neutrino masses is implemented by extending the standard model with the addition of a triplet Higgs field $\Delta$ with $Y=2$. This leads to the Yukawa coupling in the standard model which has the form:
\begin{eqnarray}
{\cal L}_Y~=~h_u \bar{Q}_LHu_R +h_d \bar{Q}_L\tilde{H}d_R+ h_\ell \bar{L}_L\tilde{H}e_R+fL^T\Delta L+h.c.\label{lagrangian}\,.
\end{eqnarray}
The Higgs potential for the model is:
\begin{eqnarray}
V(H,\Delta)~=~-\mu^2_HH^\dagger H+\lambda_H(H^\dagger H)^2+M^2_\Delta \Delta^\dagger\Delta+\lambda_\Delta (\Delta^\dagger\Delta)^2 +\\\nonumber
\lambda_{H\Delta}(H^\dagger H)(\Delta^\dagger\Delta)+\mu_\Delta HH\Delta~+~h.c.
\end{eqnarray}
with $\mu_H^2, M^2_\Delta > 0$~\footnote{Interestingly, the scalar potential of extensions of the SM with triplets has been shown to significantly impact on the perturbativity and vacuum stability of the Higgs potential. In particular the region of Higgs masses that correspond to a potential stable and perturbative  up to very high energy includes lighter masses than what predicted in the pure SM~\cite{Gogoladze:2008fj}.}.
The simultaneous presence of the couplings $f$ and $\mu_{\Delta}$ breaks lepton number and upon the EW symmetry breaking the VEV of SM doublet,  $\vev{H} = v$, triggers the VEV of $\Delta^0$ to be non-zero. We get
\begin{eqnarray}
v_\Delta \equiv \vev{\Delta^0}=\frac{\mu_\Delta v^2}{M^2_\Delta}\,.
\end{eqnarray}
{As clear from eq.~(\ref{lagrangian}) this VEV generates a mass for the neutrinos.} There are two ways to get a eV neutrino mass from this formula. Since $\mu_{\Delta}$ and $M_\Delta$ are both standard model singlet parameters, their values do not depend on the weak scale and  can have very high values  and it is natural to expect them to be of the same order, i.e.   $\mu_{\Delta}\sim M_\Delta$. Observed neutrino masses require this heavy mass to be of order $10^{14}$ GeV. This keeps the physics of $\Delta$ particles hidden from low energy experiments. On the other hand one could have $M_\Delta\sim$~TeV so that its effects are accessible not only in low-energy high-intensity experiments but also in high-energy colliders. In this case we need to tune
$\mu_{\Delta}\sim 10^{-10}$ GeV or so. It is this fine tuning of $\mu_{\Delta}$ that we want to address in this paper.

If we consider the supersymmetric version of the triplet seesaw model,  the field content of the model is: $Q, u^c, d^c , L, e^c, H_u, H_d, \bar{\Delta}, \Delta$ with a superpotential of the form:
\begin{eqnarray}
W~=~y_u {Q}H_uu^c +y_d {Q}{H}_dd^c+ y_\ell {L}{H}_de^c+fL^T\Delta L+\mu_H H_uH_d+\\ \nonumber M_\Delta \bar{\Delta}\Delta + \kappa_u \Delta H_dH_d+\kappa_d \bar{\Delta} H_uH_u\label{naivemodel}\,.
\end{eqnarray}
In this case the triplet VEV for $\Delta$ arises from the potential generated by term  $|F_{H_d}|^2$  such that $\mu_{\Delta}$ of the non-SUSY triplet model becomes $\kappa_u\, \mu_H$. Since $\mu_H$ is at least of the order of the weak scale to get a realistic EWSB, the fine tuning condition to get small neutrino masses described above translates to the extreme smallness of coupling parameter $\kappa_u\sim 10^{-10}$~\footnote{Through the mass term $M_{\Delta}$, when the VEV is generated for $\bar{\Delta}$ also $\Delta$ gets a VEV and vice-versa. 
Therefore one could get a similar conclusion for the tuning of $\kappa_{d}$ as well.}. Our goal in this paper is to extend the type II seesaw model so that this fine tuning becomes alleviated.

\subsection{Vector-like quark extension and alleviation of fine-tuning}

We extend the supersymmetric type II seesaw model of the previous section adding a heavy vector-like quark doublet with $Y=7/6$ so that its component fields, $(X_{5/3}, t')$, have charges $Q(X_{5/3})=5/3$ and $Q(t')=2/3$. We call the two vector-like doublets $Q^\prime$ and $Q^{\prime,c}$. We also add to the model a gauge singlet field $\sigma$. In order to show that this extension helps to make the TeV scale type II seesaw natural, we impose an additional $\mathbb{Z}_4$ symmetry on the model under which $\Delta, \bar{\Delta}, Q^{\prime c}, \sigma \to -\Delta, -\bar{\Delta}, -Q^{\prime c}, -\sigma$, $e^{c}\to -i e^{c}$, and $L\to iL$ and all other fields are even. In Table~\ref{charges} we summarize the particle content beyond the MSSM and the charges under the SM gauge group and under the $\mathbb{Z}_{4}$ symmetry. We find the following new Yukawa  superpotential:
\begin{eqnarray}
\label{susymodel}
{\cal W}^\prime~=~y_u {Q}H_uu^c +y_d {Q}{H}_dd^c+ y_\ell {L}{H}_de^c+fL^T\Delta L+\mu_H H_uH_d+\\ \nonumber M_\Delta \bar{\Delta}\Delta+f_q{Q'}^c\Delta Q+ \lambda 
Q^\prime H_d u^c+ y_{Q^{'}} {Q}^{\prime,c} Q^{\prime}\sigma\,. \end{eqnarray}

\begin{figure}[t]
\begin{center}  
\includegraphics[width=0.5 \linewidth]{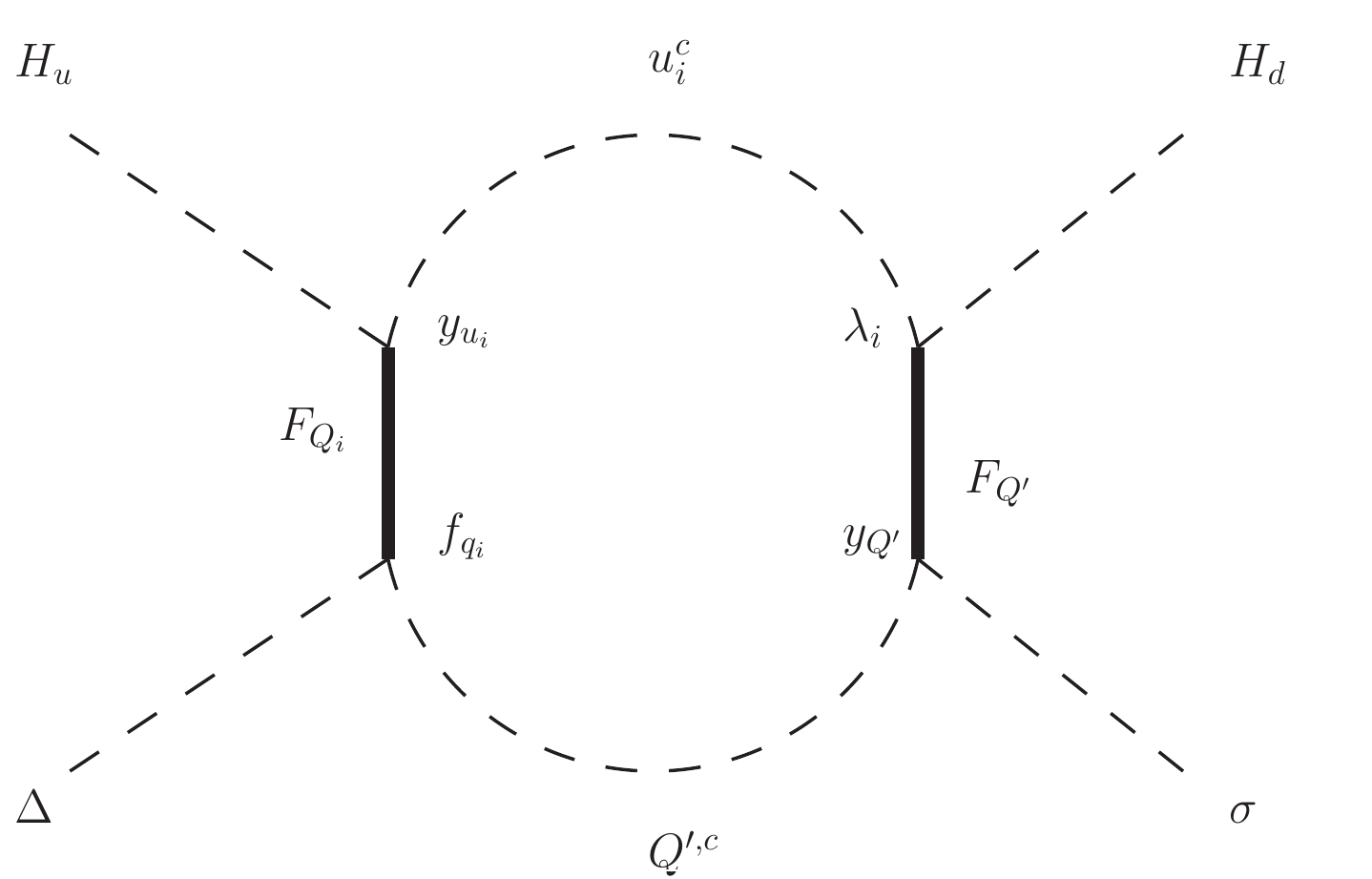}
\end{center}
  \caption{One loop diagram that induces finite $\Delta H_uH^*_d\sigma^*$ vertex.\label{oneloop}}
\end{figure}

This superpotential is remarkable in that it has interactions of the heavy quarks with several states of the Higgs sector, in particular we remark the interactions with the triplet $\Delta$ and with the Higgs doublet $H_{d}$. {In principle our symmetry allows in the superpotential  R-parity and baryon number violating terms $u^{c}d^{c}d^{c}$, which are expected from the fact that our $\mathbb{Z}_{4}$ symmetry is equivalent to lepton number when SM states are considered. In the following we do not consider these type of interactions, assuming for simplicity that they are forbidden by a suitable symmetry.}
Furthermore we note that from our symmetry follows the absence of the terms $\kappa_u \Delta H_dH_d+\kappa_d \bar{\Delta} H_uH_u$ from the superpotential. As a result, at the tree level, there no induced VEV for the $\Delta$ field. It arises at the one loop level after SUSY breaking from the diagram in Fig.~\ref{oneloop}.
{This diagram is induced by 
\begin{eqnarray}
F_{Q^{\prime}} &\supset & \lambda H_{d} u^{c} + y_{Q^{'}} Q^{\prime,c}\sigma \,, \\
F_{Q}&\supset& f_{q}Q^{\prime,c}\Delta + y_{u} H_{u}u^{c}\,,
\end{eqnarray}
and generates  a finite 
and calculable coupling $\Delta H_u^{*}H_d\sigma^*$ 
which gives rise to an effective $\mu_{\Delta,{\rm eff}}$ when the singlet field $\sigma$ gets a VEV

\begin{equation}
\mu_{\Delta,{\rm eff}}=\frac{y_{Q^{'}}v_{\sigma}}{16\pi^{2}}\sum_{i}y_{u_{i}}f_{q_{i}}\lambda_{i} \equiv \frac{y_{Q^{'}}v_{\sigma}}{16\pi^{2}} \yeff \,.\label{mudeltasum}
\end{equation}

\begin{table}[t]
\begin{center}
\begin{tabular}{c|c|c}
 & $G_{SM}$ & $\mathbb{Z}_{4}$\\
\hline 
\hline ~& &\\
 $\Delta=\begin{pmatrix}\Delta^{++}\\
\Delta^{+}\\
\Delta^{0}
\end{pmatrix}$ & $(\boldsymbol{1},\boldsymbol{3},1)$ & $-1$\tabularnewline
 &  & \tabularnewline
$\bar{\Delta}$ & $(\boldsymbol{1},\boldsymbol{3},-1)$ & $-1$\tabularnewline
 &  & \tabularnewline
$Q^{\prime}=\begin{pmatrix}X_{5/3}\\
t^{\prime}
\end{pmatrix}$ & $(\boldsymbol{3},\boldsymbol{2},\frac{7}{6})$ & 1\tabularnewline
 &  & \tabularnewline
$Q^{\prime,c}$ & $(\boldsymbol{3^{*}},\boldsymbol{2^{*}},-\frac{7}{6})$ & $-1$\tabularnewline
 &  & \tabularnewline
$\sigma$ & $(\boldsymbol{1},\boldsymbol{1},0)$ & $-1$\tabularnewline
 &  & \tabularnewline
$L$ & $(\boldsymbol{1},\boldsymbol{2},-\frac{1}{2})$ & $i$\tabularnewline
 &  & \tabularnewline
$e^{c}$ & $(\boldsymbol{1},\boldsymbol{1},1)$ & $-i$\tabularnewline
\end{tabular}
\end{center}
\caption{\footnotesize Transformation properties of the field content beyond the MSSM and of the MSSM fields charged under the $\mathbb{Z}_{4}$ symmetry of our model. \label{charges}}
\end{table}

All squark flavors run into the loop, therefore we define $\yeff=\sum_{i}y_{u_{i}}f_{q_{i}}\lambda_{i}$ which represents the combination of flavored new Yukawa couplings $\lambda$, $f_{q}$ and SM Yukawa couplings $y_{u}$ that is relevant for the generation of neutrino masses. For not  hierarchical $\lambda_{i}$ and $f_{q_{i}}$ one can assume that the above formula is dominated by the contribution of the third generation 
\begin{equation}
\mu_{\Delta,{\rm eff}}\simeq \frac{y_{Q^{'}}v_{\sigma}}{16\pi^{2}} y_{t}f_{q_{3}}\lambda_{3} \,.\label{mudeltathird}
\end{equation}
Including this term in the minimization of the potential induces a non-zero VEV for $\Delta^0$ that in general is given by
\begin{eqnarray}
v_\Delta = \mu_{\Delta,{\rm eff}} \frac{ v_u v_d}{M^2_\Delta}\,,
\end{eqnarray}
which for $\yeff\simeq y_{t}\lambda_{3}f_{q_{3}}$ reduces to
\begin{eqnarray}
v_\Delta \simeq \frac{f_{q_{3}}\lambda_{3}}{16\pi^{2}} M_{Q'} \cos\beta \frac{ m_{t} v}{M^2_\Delta}\,,
\end{eqnarray}
where $\tan\beta=\vev{H_{u}}/\vev{H_{d}}$, and $M_{Q'}=y_{Q'}v_{\sigma}$ is the mass of the exotic quark generated by the VEV of $\sigma$.
For $M_{Q'}\sim M_{\Delta}\sim \TeV$ at large $\tan\beta$ and considering only the contribution to neutrino masses from the third generation this corresponds to 
\begin{equation} v_{\Delta}\simeq  \left(\frac{\lambda_{3} f_{q_{3}}}{ 10^{-7}}\right)\left(\frac{20}{\tan\beta}\right)\left(\frac{1\tev}{M_{\Delta}}\right)^{2}\left(\frac{M_{Q^{'}}}{1 \tev}\right)~\ev\label{mneutrino}\,.
\end{equation}

Let us now recall that current neutrino data for a normal hierarchy imply that the triplet coupling matrix $f_{ij}$ has the form~\cite{Dev:2013uq}:
\begin{eqnarray} 
f~=\frac{10^{-2}}{v_{\Delta}/\textrm{eV}}\left(\begin{array}{ccc} 0.31 - 0.12i &-0.09 + 0.32i &-0.72 + 0.37i \\ -0.09 + 0.32i & 2.53 + 0.04i&	2.19 + 0.01i \\ 
	-0.72 + 0.37i	&2.19 + 0.01i	&3.07 \end{array}\right)\,.
\end{eqnarray}
Constraints of the flavor violating process $\mu\to 3e$ then imply that $f_{ee}f_{e\mu}\leq 4\times 10^{-5}$ for TeV $M_{\Delta^{++}}$. In turn this implies that $v_\Delta \geq 0.5$ eV. Looking at eq.~(\ref{mneutrino}), we find that
 to obtain a realistic mass scale for the neutrino mass generated by the type-II seesaw it is no longer necessary to tune a single coupling to be very small, as it is the case for the generic type-II seesaw model of eq.~(\ref{naivemodel}). In fact one can attain the right size of neutrino masses by taking the product of two couplings $\lambda_3\cdot  f_{q_{3}}\geq 10^{-7}$. This alleviates significantly the fine tuning problem.
%

\subsection{Quark masses and mixings to $X_{5/3}$ and $t'$\label{sec:mixing}}
After symmetry breaking, the $X_{5/3}$ picks up mass of order $\vev{\sigma}$. Guided by the idea that this model describes physics a the weak scale, we choose $\vev{\sigma}\simeq 500-1000$ GeV. {The top partner $t'$ also picks up the same mass, however it mixes with SM up-type quarks $u_{i}$ due to the $\lambda_{i}$ and $f_{q_{i}}$ couplings and leads to the following mass matrix in the $u_{i}-t'$ basis:
\begin{eqnarray}
M_{u_{i}-t'}=\left(\begin{array}{cc} y_{u_{i}} v_u & f_{q_{i}} v_\Delta\\ \lambda_{i} v_d & \vev{\sigma}\end{array}\right)\,.
\end{eqnarray}
Due to the smallness of the triplet VEV, the mixing of $t'_L$ to $u_{i}$ in general dominates over the mixing of $t'_{R}$. The mixing of $t'_L$ to $u_{i}$ is given by $\theta_{u_{i}t'}\equiv\lambda_{i} v_d/\vev{\sigma}\sim 10^{-3}$.  We remark that the smallness of this mixing is controlled by the requirement of getting small neutrino masses.} Without this restriction, this mixing could be quite significant - although still less than 10-30\% or so depending on what the mass of the $X_{5/3}$ quark is.  From this mass matrix, we also learn that the splitting between $X_{5/3}$ and $t'$ is very small (less than a GeV). The rest of the quarks and leptons get their masses as in the standard model. The only effect on the standard model quark sector is that there will very small mixings between each quark and the $t'$.

{Besides introducing a mixing between new matter and SM matter, the coupling $\lambda$ has its most apparent effect in the modification of the decay of the heavy quarks with respect to the minimal models of heavy quarks, where they usually decay to Goldstone bosons in the form of longitudinal SM gauge bosons. 
As remarked above, in principle also $f_{q}$ mediates a mixing, but it is significantly suppressed by the smallness of the VEV $v_{\Delta}$. We will see below that the interactions mediated by  $f_q$, as well as the one mediated by $\lambda$, have several interesting phenomenological implications for the decay of the new fields of our model. In particular they  induce new decay modes for the heavy quarks, making them phenomenologically distinct from the heavy quarks of minimal models.
}

\section{Decays of the $\frac{5}{3}$ charge quark and the seesaw triplet\label{lhc}}
The states $\Delta$ and $Q^{'}$ have a peculiar phenomenology, especially the $X_{5/3}$ and the $\Delta^{++}$ states. In fact their exotic electric charge gives rise to specific signatures characterized for instance by the presence of leptons of same sign that can also have a resonant structure. 
These features are indeed exploited in theoretical and experimental studies on the observation of these states~\cite{q5/3,typeii,Hektor:2007uu,Chatrchyan:2012af,ATLAS:2012qe,Chatrchyan:2012vu,q5/3cms}.

Despite the large literature about the states $\Delta$ and $Q^{'}$ the phenomenology of models where both these states appear is less studied. We have seen above how the simultaneous presence of these states can induce neutrino masses radiatively. For this goal are necessary several interactions beyond those of the SM and the MSSM. In particular from \eq{mneutrino} we see that three couplings of the model of \eq{susymodel} are involved in the neutrino mass generation:
\begin{itemize}
\item $\lambda$ which mediates interactions among $Q^{'}$, $u^{c}$ and $H_{d}$ and that after EWSB it also causes the $t'-u_{i}$ mixing;
\item $f$ which gives interactions among lepton doublets and the Higgs triplet $\Delta$;
\item $f_{q}$ which gives  interactions among $Q^{'}$, $Q$ and the Higgs triplet $\Delta$.
\end{itemize}

Only the product of these couplings is constrained by neutrino masses, $\lambda f f_{q}\simeq 10^{-7}$, so that there is still significant freedom for the value of each single coupling.
The masses of the new states $Q^{'}$ and $\Delta$ are expected to be of the same order and there is no substantial preference for one being heavier than the other~\footnote{In fact one can even find  discrete symmetries slightly different from our $\mathbb{Z}_{4}$ that would give both the mass of $\Delta$ and that of $Q^{'}$ from renormalizable Yukawa superpotential interactions. In that case the resulting mass for the two states is controlled by two couplings with no a priori hierarchy.}. We first consider the case $M_{X_{5/3}}>M_{t'}>M_{\Delta}$, which is somewhat more interesting than the converse because in this case scalar non-colored states can appear in the decay chain of copiously produced colored objects. In this case the exotic $X_{5/3}$ has several possible decay modes discussed in the following.

Weak gauge interactions mediate decays into states of the $Q^{'}$ multiplet in the same way they do in  typical models where heavy quarks appear:
\begin{eqnarray}
X_{5/3}&\to& W^{+} t'\,,\\
X_{5/3}&\to& W^{+} t\,,\label{gaugemodes}
\end{eqnarray}
where the latter involves a $t-t'$ mixing. {The first process is kinematically forbidden for most of the parameter space of our model or is generically suppressed for virtual $W$ by the small mass difference between $X_{5/3}$ and $t'$ or by the suppression in the $t-t'$ mixing from the small $v_{d}/\vev{\sigma}$ that we discussed in Section~\ref{sec:mixing}}.

Our model is characterized by the existence of decay modes mediated by $\lambda$
\begin{eqnarray} 
t' &\to& H_{d}^{0} \, u_{i}\,,\\
X_{5/3} &\to& H_{d}^{+}\, u_{i} \,,
\end{eqnarray}
where $H_{d}^{+}$($H_{d}^{0}$) is a component of the charged (neutral) Higgs boson that is mostly made out of the doublets~\footnote{In this model the VEV of the triplet field is small and the trilinear terms of the type $\Delta HH$ are forbidden at tree level, therefore it is rather natural to assume that the triplets and the doublets do not mix significantly.}.

Interestingly, these modes can be the dominant ones if one takes $\lambda$ to be the largest coupling and ascribes the smallness of neutrino masses to $f\cdot f_{q}$. 
In fact we estimate that the coupling $\lambda$ could be of order 0.1 without causing tensions in precision observables in the $Z\to b \bar{b},\,Z\to hadrons$ decays, and $D$ meson mixing~\cite{Isidori:2010zr}, which are potentially sensitive to this coupling~\footnote{Ultimately these effects are small thanks to the smallness of the $u_{i}-t'$ mixing described in Section~\ref{sec:mixing} and the mass of the $Q'$ fields in the TeV range. It should also be noted that this theory predominantly generates  chirality conserving operators for the FCNC processes. See a more extended discussion in Sec.~\ref{other}.}.
Depending on the way the states $H^{+,0}_{d}$  decay, these decay modes could be different enough from the usual decay modes assumed in searches for heavy quarks and they may even cause a significant relaxation of the bounds on the existence of $Q^{'}$.

 In the triplet-doublets decoupling limit, and for moderate or large $\tan\beta=\vev{H_{u}^{0}}/\vev{H_{d}^{0}}$, the field $H_{d}^{0}$ is very close to the heavy mass eigenstates of the MSSM scalar spectrum $H,A$. Therefore the dominant $t'$ decays mediated by $\lambda$ are expected to be
\begin{eqnarray}
t' &\to& H \,u_{i}\,, \label{tprimelambdadecayH}\\
t' &\to& A \,u_{i}\,, \label{tprimelambdadecayA}
\end{eqnarray}
whereas $t' \to h\, u_{i}$ is likely to be sub-dominant due to the value of $\tan\beta$. 
In principle the decay rate may depend on the flavor of the up-type quark involved in eq.(\ref{tprimelambdadecayH}) and (\ref{tprimelambdadecayA}).
We expect $$H,A\to b\bar{b}$$ to be among the dominant decay modes, and likely the dominant decay mode, of the heavy Higgs bosons. In fact decays into Higgs bosons and gauge vectors are typically suppressed by phase-space or $\tan\beta$. The decay in fermionic partners of the electroweak gauge bosons  
$$H\to\chi^{0}\chi^{0},\chi^{+}\chi^{-}$$
are in principle possible, although the decay rate is highly dependent on mass and mixing parameters of the neutralino-chargino sector. In any case it is worth stressing the possibility. In fact a significant branching fraction in this decay mode  would result in a significant change of the observable signature of the production of the heavy quarks.

Sticking to the limit where the triplet is decoupled from the doublets, the $X_{5/3}$ decays 
\begin{eqnarray}
X_{5/3} &\to& H^{+} u_{i}\to t\,\bar{b} \,u_{i} \label{lambdadecayX}
\end{eqnarray}
where the decay of the charged Higgs boson into fermions is favored by the smallness of $v_{d}$ and the decay into $t\bar{b}$ is assumed to be kinematically allowed~\footnote{Although it is not strictly forbidden to have lighter charged Higgs bosons, we focus on the case of a charged Higgs heavier than 350 GeV such to straightforwardly satisfy the  limits on the $b\to s\gamma$ decay rate~\cite{The-BABAR-Collaboration:2012qf} and $t\to H^{+}b$ searches~\cite{CMS-PAS-HIG-11-008}.}. The hierarchy between $v_{u}$ and $v_{d}$ suppresses also other decays such as $X_{5/3} \to W^{+}\,u_{i}$, making \eq{lambdadecayX} the main decay mode of $X_{5/3}$ when $\lambda$ dominates.

We see that having two Higgs doublets plays an essential role in the phenomenology of the exotic fermions that can now decay in the extra physical degrees of freedom of the scalar sector (see Refs.~\cite{Vecchi:2013kx,Kearney:2013fj,Kearney:2013kx} for a recent related discussion about heavy quarks phenomenology in the context of models of strongly interacting EWSB). The different decay pattern of these heavy quarks is likely to make the standard searches less sensitive to the type of object present in this model.
However,  we do not expect this scenario to be free from experimental bounds. In fact searches for resonances that decay in $tb$ final state~\cite{CMS-PAS-B2G-12-010,Abazov:2006aj,Acosta:2002nu}  are potentially sensitive to the production of $H^{+}$ in cascades of the exotic quarks as in  the decay \eq{lambdadecayX}. Furthermore the searches for heavy Higgs bosons of the MSSM \cite{CMS-Collaboration:2013fr,Chatrchyan:2011nx} are in principle sensitive to the decays of heavy quarks into states of the extended Higgs sector. Also, an extension of searches for top quarks in association with Higgs bosons of some more general scope of the current ones~\cite{CMS-Collaboration:2013ys} would certainly allow to investigate in full generality the phenomenology of heavy quarks.

\begin{table}[h]
\begin{raggedright}
\begin{center}
\begin{tabular}{cccc}
\toprule 
\multicolumn{2}{c}{Decay} & Rate & Dominant for\tabularnewline
\midrule
\midrule 
\multicolumn{4}{c}{$M_{Q'}>M_{\Delta}$}\tabularnewline
\midrule
\midrule  
\multicolumn{2}{c}{$Q'_{\overrightarrow{\,\,\lambda\,\,}}H_{d}u^{c}$} & unsuppressed & $\lambda\gg f_{q}$\tabularnewline\addlinespace[3mm]
\midrule 
\multicolumn{2}{c}{$Q'_{\overrightarrow{\,\, f_{q}\,\,}}\Delta q{}_{\overrightarrow{\,\, f\,\,}}LLq$} & unsuppressed & $f_{q}\gg\lambda$\tabularnewline\addlinespace[3mm]
\midrule 
\addlinespace[3mm]
\multicolumn{2}{c}{$Q'_{\overrightarrow{\,\, g\,\,}}V u^{c}$ } & suppressed by $\frac{\lambda v_{d}}{\left\langle \sigma\right\rangle }$ & $\lambda\gg f_{q}$ and $m_{H_{d}}>m_{Q'}$\tabularnewline\addlinespace[3mm]
\midrule 
\midrule
\multicolumn{4}{c}{$M_{\Delta}>M_{Q'}$}\tabularnewline
\midrule 
\midrule
\multirow{3}{*}{$\Delta_{\overrightarrow{\,\, f_{q}\,\,}}Q'q_{L}$$\begin{cases}
\\
\\
\\
\\
\\
\end{cases}$} & $Q'_{\overrightarrow{\,\,\lambda\,\,}}H_{d}u^{c}$ & unsuppressed & $f_{q},\lambda\gg f$\tabularnewline\addlinespace[3mm]
\cmidrule{2-4} 
 & $Q'_{\overrightarrow{\,\, f_{q}\,\,}}q_{L}\Delta_{\overrightarrow{\,\, f\,\,}}^{*}q_{L}LL$ & 3-body suppressed 
 & $f_{q}\gg\lambda,f$\tabularnewline\addlinespace[3mm]
\cmidrule{2-4} 
 & $Q'_{\overrightarrow{\,\, g\,\,}}Vu^{c}$  & suppressed by $\frac{\lambda v_{d}}{\left\langle \sigma\right\rangle }$ & $\lambda\gg f_{q}\gg f$ 
 and $m_{H_{d}}>m_{Q'}$
 \tabularnewline\addlinespace[3mm]
\midrule 
$\Delta_{\overrightarrow{\,\, f\,\,}}LL$ &  & unsuppressed & $f\gg\lambda,f_{q}$\tabularnewline
\bottomrule
\end{tabular}
\end{center}
\end{raggedright}

\caption{\footnotesize  \label{summarydecay} Summary of the decay patterns of
the non-SUSY particles of our model. The shorthands
$L=\ell,\nu$, and $V=Z,W$  must be spelled out in the suitable
combination for each final state. For the decay of $H_{d}$ in the
large $\tan\beta$ limit and neglecting the small doublet-triplet
mixing the decays $H_{d}\to f\bar{f}$ or $\chi\bar{\chi}$ are both possible, but the $f\bar{f}$ final state is generically favored.}
\end{table}

The coupling $f_{q}$ mediates several  decays of $Q^{'}$ that are distinctive of our model. We estimate that a coupling of order order 0.1 would respect available constraints from data meson oscillation~\cite{Isidori:2010zr} and $Z$ pole precision observables~\cite{Beringer:1900zz}. Therefore we can consider a scenario where the decay of $Q^{'}$ is dominated by the coupling $f_{q}$ while the smallness of the neutrino mass is ascribed to the couplings $\lambda$ and $f$. The coupling $f_{q}$ mediates
\begin{eqnarray}
X_{5/3}&\to& \Delta^{++} \, d_{i}  \label{Xdeltad}\,,\\
X_{5/3}&\to& \Delta^{+} \, u_{i} \label{Xdeltau}\,,
\end{eqnarray}
with strength that in principle may depends on the flavor of the down-type and up-type quarks $d_{i}$ and $u_{i}$. The mode of \eq{Xdeltad} is new and distinctive of the present model with respect to the usual models where the state $Q^{'}$ is not involved in neutrino mass generation and therefore it does not have the interaction $f_{q} Q^{'}\Delta Q$. The way this decay can be detected in practice depends a lot on the decay of the $\Delta$, which we discuss below.
The mode \eq{Xdeltau} and other decay modes of $t'$ mediated by $f_{q}$ such as 
\begin{eqnarray}
t' &\to& \Delta^{+} \, d_{i}\,,\nonumber \\
t' &\to& \Delta^{0} \, u_{i}\,,\label{stdlikemodes}
\end{eqnarray}
are new as well. However, as they involve Higgs states of charge one or zero, they are likely to result in final states quite similar to those of $t'\to Z t,\,ht$ and $X_{5/3}\to W^{+}t$ and so they are less distinctive of this model. In case of observation of an excess in the searches sensitive to these decay modes one should take into account the possibility of extra contributions, with possibly peculiar flavor structure, that contribute to the lepton rich final states of these BSM searches~\cite{Chatrchyan:2012af,ATLAS:2012qe,Chatrchyan:2012vu,q5/3cms}. Due to the similarities among the standard decay modes considered in the searches of heavy quarks and the decays \eq{stdlikemodes} we can take the current bounds from~\cite{Chatrchyan:2012af,ATLAS:2012qe,Chatrchyan:2012vu,q5/3cms} and future updates as a good estimate of the mass bound on $m_{Q'}$ for the scenario where $f_{q}$ dominates the decay of $Q'$.

The triplet $\Delta$ decays are quite well studied in the context of type II seesaw model, however the picture that emerges in this model can be substantially different. 
In a way similar to the standard type II seesaw model the gauge interactions mediate
\begin{eqnarray}
\Delta^{--}&\to& W^{-}W^{-} \,,\\
\Delta^{--}&\to& W^{-}\Delta^{-}\,,
\end{eqnarray}
that are suppressed by the VEV $\vev{\Delta^{0}}$ and by the smallness of the typical splitting of the components of the $\Delta$ triplet. Therefore these interactions do not typically have impact of the phenomenology of $\Delta$ and we neglect them~\footnote{It should be remarked that being a SUSY model there are two Higgs doublets, and therefore more physical degrees of freedom into which the $\Delta$ can decay. In fact, in addition to the listed modes there could be interactions with the additional scalars $H^{\pm},H,A$ of the MSSM. For the decay the most important interactions should arise from trilinear terms. These are either originating from quartic term of the form $\Delta^{2} H^{2}$ or from genuine cubic terms $\Delta HH$. In both cases these interactions are suppressed. In fact the quartic terms affect the decay of $\Delta$ only after one of the $\Delta^{0}$ has taken a VEV, which we know must be a small quantity in this model. On the other hand trilinear terms of the form $\Delta HH$ are forbidden  at tree-level in this model, hence do not in general change the phenomenology of the decay of $\Delta$.}.

As in the standard type II seesaw model, the interaction $f$ mediates the decays into leptons that are used in the current searches for the seesaw triplet~\cite{Collaboration:2012uq}
\begin{eqnarray}
\Delta^{--}&\to& \ell \ell\,, \label{deltafmodesll}\\
\Delta^{-} &\to& \ell \bar{\nu}\,, \label{deltafmodeslnu} \\
\Delta^{0} &\to& \nu \bar{\nu}\,. \label{deltafmodesnunu}
\end{eqnarray}
When the decay of $\Delta$ is dominated by $f$ the searches for doubly charged resonances~\cite{Collaboration:2012uq} are in force. Depending on the lepton flavor combination that appear in \eq{deltafmodesll} and \eq{deltafmodeslnu} a bound on the productions cross-section varying from about 1~fb to about 100~fb is in force. Considering that the source of the doubly charged scalar in this model would be the colored fermion $X_{5/3}$ this translates into a bound on the mass of the $Q^{'}$ up to 1 TeV.
Therefore in this model the bounds from the peculiar decay mode \eq{Xdeltad} are stronger than the bounds on heavy quark from searches that look for the  signatures of minimal models of heavy quarks~\cite{Chatrchyan:2012af,ATLAS:2012qe,Chatrchyan:2012vu,q5/3cms}.

Differently from the standard type II seesaw model without the $Q^{'}$ fields, in our model  $\Delta$ can decay via the interactions mediated by the couplings $\lambda$ and $f_{q}$. For instance the  doubly charged scalar has the decay mode  
\begin{eqnarray}
\Delta^{++}&\to& \bar{d}_{i} X_{5/3}^{*} \to \bar{d}_{i } u_{j}H_{d}^{+}  \to \bar{d}_{i }  u_{j} t\bar{b}\label{deltathreebody}
\end{eqnarray}
which involves the product of couplings $\lambda f_{q}$. This mode can dominate when the coupling $f$ is by far the smallest among $\lambda,f_{q},f$. A very small $f$ may be suggestive of some tuning of the couplings of the model, going against the original motivation for our model. However the resulting phenomenology of \eq{deltathreebody}, is quite different from the case where $f$ dominates the decay of $\Delta$, and therefore we believe it deserves some attention. Furthermore, especially for light $\Delta$, the coupling $f$ may be required to be small to comply with lepton number violation bounds  from $\mu\to eee$ and radiative processes such as $\mu\to e\gamma$~\cite{Hundi:2012rt}.

\bigskip

For the case where the triplet is heavier than the exotic quark, $M_{\Delta}>M_{Q'}$, there are significant differences. Most importantly the production of the electroweak states of $\Delta$ cannot be enhanced by the production of $\Delta$ in decay chains of colored states. However one can still have the doublet-like states $H_{d,u}$ in such decay chains, which is still a interesting signature of this type of heavy quarks.

In the case $f\gg\lambda,f_{q}$ the triplet $\Delta$ decays as in the usual type-II seesaw model. In particular there is a decay $$\Delta^{--}\to\ell\ell$$ and the bounds from~\cite{Collaboration:2012uq} are in force.

In the case $f\ll\lambda,f_{q}$ the triplet cannot decay into leptons and it has to decay through the coupling $f_{q}$ into a left-handed SM quark and a heavy quark $Q'$
\begin{equation}
\Delta\to Q'\, q_{L,SM}\,.
\end{equation}
Therefore the decay of $\Delta$ is effectively giving rise to an extra source of $Q'$. This extra source of $Q'$ is certainly sub-dominant to the QCD production of $Q'$. However we can exploit this fact to use the phenomenology of $Q'$  as a single probe to investigate both the existence of $Q'$ and $\Delta$. 

After the production, either through QCD gauge interactions on in the decay of $\Delta$, the decay of $Q'$ can either be ruled by $\lambda$, $f_{q}$ or gauge interactions.
The decay through the coupling $\lambda$ is the only one that can go through without suppression. In fact it gives a two-bodies decay $Q'\to H_{d} u^{c}$ that results in signals similar to \eqs{tprimelambdadecayH},(\ref{tprimelambdadecayA}),(\ref{lambdadecayX}) discussed above.
These decays, when kinematically allowed, are expected to be dominant because the decays mediated by $f_{q}$ or gauge interactions pay either a phase-space or a mixing suppression. As stressed already in the discussion of the case $M_{Q'}>M_{\Delta}$ the decay modes \eqs{tprimelambdadecayH},(\ref{tprimelambdadecayA}),(\ref{lambdadecayX}) give rise to interesting signatures that characterize our type of non-minimal heavy quarks. Bounds on these signatures can be obtained recasting (or extending) current searches as detailed above.

The decay meditated by $f_{q}$ gives a 3-bodies decay into an off-shell $\Delta$. The virtual $\Delta$ can decay only into leptons, that are the only light final state available and overall we have a decay of $Q'$ {
\begin{equation}
Q' \to \Delta^{*}q_{L,SM}\to \ell\ell\, q_{L,SM}\,.\label{fqthreebody}
\end{equation}}
This signature is in general very similar to the one with non-resonant same sign and opposite sign leptons investigated in the searches for minimal heavy quarks~\cite{Chatrchyan:2012af,ATLAS:2012qe,Chatrchyan:2012vu,q5/3cms}. Therefore we expect that, up to differences in the efficiencies, these searches would constrain the scenarios where $Q'$ decays in three-bodies via the coupling $f_{q}$. This decay mode is expected to dominate when either the states $H,A,H^{+}$ are too heavy for the decays eqs.(\ref{tprimelambdadecayH}),(\ref{tprimelambdadecayA}),(\ref{lambdadecayX}) to be on-shell, or there is a large hierarchy of couplings $f_{q} > 16 \pi^{2} \lambda$.

The 3-bodies decay mediated by $f_{q}$ has to compete with the gauge decay modes, which are the same as in \eq{gaugemodes}. In this case there is a suppression of order $v_{d}/\vev{\sigma}$ from the  $u_{i}-t'$ mixing angle, which is of the same size of the 3-body phase space suppression. Therefore the dominance of the gauge modes with respect to the 3-bodies decays of \eq{fqthreebody} depends on the detailed hierarchy among the couplings of the model.
In any case, whether are the gauge modes or the 3-bodies modes of \eq{fqthreebody} that dominates the decay, the standard searches~\cite{Chatrchyan:2012af,ATLAS:2012qe,Chatrchyan:2012vu,q5/3cms} should apply up to minor changes in the mass reach.

A summary of the decay of non-SUSY states is presented in Table~\ref{summarydecay}.

\subsection{SUSY modes}
\begin{table}
\begin{raggedright}
\begin{center}
\begin{tabular}{cccc}
\toprule 
\multicolumn{2}{c}{Decay} & Rate & Dominant for\tabularnewline
\midrule
\midrule 
\multicolumn{4}{c}{$M_{Q'}>M_{\Delta}$}\tabularnewline
\midrule
\midrule 
\multicolumn{2}{c}{$\tilde{Q}'_{\overrightarrow{\,\,\lambda\,\,}}\tilde{H}_{d}u^{c}$} & unsuppressed & $\lambda\gg f_{q}$\tabularnewline\addlinespace[3mm]
\midrule 
\multicolumn{2}{c}{$\tilde{Q}'_{\overrightarrow{\,\,\lambda\,\,}}H_{d}\tilde{u}^{c}$} & unsuppressed & $\lambda\gg f_{q}$\tabularnewline\addlinespace[3mm]
\midrule 
\multicolumn{2}{c}{$\tilde{Q}'_{\overrightarrow{\,\, f_{q}\,\,}}\tilde{q}_{L}\Delta_{\overrightarrow{\,\, f\,\,}}$
leptons + jets + mET} & unsuppressed & $f_{q}\gg\lambda$\tabularnewline\addlinespace[3mm]
\midrule 
\multicolumn{2}{c}{$\tilde{Q}'_{\overrightarrow{\,\, f_{q}\,\,}}q_{L}\tilde{\Delta}_{\overrightarrow{\,\, f\,\,}}$
leptons + jets + mET} & unsuppressed & $f_{q}\gg\lambda$\tabularnewline\addlinespace[3mm]
\midrule 
\addlinespace[3mm]
\multicolumn{2}{c}{$\tilde{Q}'_{\overrightarrow{\,\, g\,\,}}V\tilde{u}^{c}$ and $\tilde{Q}'_{\overrightarrow{\,\, g\,\,}}\tilde{\lambda}_{i}u^{c}$ } & \multirow{1}{*}{suppressed by $\frac{\lambda v_{d}}{\left\langle \sigma\right\rangle }$} & \multirow{1}{*}{$\lambda\gg f_{q}$~$^{a}$ 
}\tabularnewline\addlinespace[3mm]
\midrule
\midrule 
\multicolumn{4}{c}{$M_{\Delta}>M_{Q'}$}\tabularnewline
\midrule 
\midrule
\multirow{6}{*}{$\tilde{\Delta}_{\overrightarrow{\,\, f_{q}\,\,}}\tilde{Q}'q_{L}$$\begin{cases}
\\
\\
\\
\\
\\
\\
\\
\\
\\
\\
\end{cases}$} & $\tilde{Q}'_{\overrightarrow{\,\,\lambda\,\,}}\tilde{H}_{d}u^{c}$ & unsuppressed & $f_{q},\lambda\gg f$\tabularnewline\addlinespace[3mm]
\cmidrule{2-4} 
 & $\tilde{Q}'_{\overrightarrow{\,\,\lambda\,\,}}H_{d}\tilde{u}^{c}$ & unsuppressed & $f_{q},\lambda\gg f$\tabularnewline\addlinespace[3mm]
\cmidrule{2-4} 
 & $\tilde{Q}'_{\overrightarrow{\,\, f_{q}\,\,}}\tilde{q}_{L}\Delta_{\overrightarrow{\,\, f\,\,}}^{*}$
leptons + jets + mET & 3-body suppressed & $f_{q}\gg\lambda,f$\tabularnewline\addlinespace[3mm]
\cmidrule{2-4} 
 & $\tilde{Q}'_{\overrightarrow{\,\, f_{q}\,\,}}q_{L}\tilde{\Delta}_{\overrightarrow{\,\, f\,\,}}^{*}$
leptons + jets + mET & 3-body suppressed & $f_{q}\gg\lambda,f$\tabularnewline\addlinespace[3mm]
\cmidrule{2-4} 
 & $\tilde{Q}'_{\overrightarrow{\,\, g\,\,}}V\tilde{u}^{c}$  & suppressed by $\frac{\lambda v_{d}}{\left\langle \sigma\right\rangle }$ &  $\lambda\gg f_{q}\gg f$~$^{a}$
 \tabularnewline\addlinespace[3mm]
\cmidrule{2-4} 
 & $\tilde{Q}'_{\overrightarrow{\,\, g\,\,}}\chi u^{c}$  & suppressed by $\frac{\lambda v_{d}}{\left\langle \sigma\right\rangle }$ & $\lambda\gg f_{q}\gg f$~$^{a}$
\tabularnewline\addlinespace[3mm]
\midrule 
$\tilde{\Delta}_{\overrightarrow{\,\, f,g\,\,}}\tilde{L}L,W\tilde{\Delta^{'}}$ & leptons + mET & unsuppressed & $f\gg\lambda,f_{q}$\tabularnewline
\bottomrule

\end{tabular}
\end{center}
\par\end{raggedright}

\caption{\footnotesize \label{tab:summarydecaysusy} Summary of the decay patterns of the
SUSY particles of our model. The shorthands $L=\ell,\nu$,
and $V=Z,W$  must be spelled out in the suitable combination
for each final state. For the decay of $H_{d}$ in the large $\tan\beta$
limit and neglecting the small doublet-triplet mixing the decays $H_{d}\to f\bar{f}$
and $\chi\bar{\chi}$ are both possible, however the decay into $f\bar{f}$ is generically favored. The decays of the SUSY states higgsino $\tilde{H}_{d}$
and gaugino $\tilde{\lambda}_{i}$ are more model dependent and we
leave them unspecified here.\\
{\scriptsize $^{\it a}$ For this mode to dominate $m_{H_{d}}>m_{Q'}$ is required, otherwise the $H_{d}^{0}$ modes are kinematically allowed and unsuppressed.}}
\end{table}
Given the presence of SUSY partners we consider the phenomenology of these particles in our model, focusing mostly on the new features with respect to the MSSM and SUSY seesaw model. We have a SUSY partner of the seesaw triplet, that we denote as tripletino $\tilde{\Delta}$, and scalar partners of the heavy quarks, the scalar heavy quarks $\tilde{Q}^{\prime}$ and $\tilde{Q}^{\prime,c}$.

When the coupling $f$ dominates among the new couplings, our model is essentially reproducing the phenomenology of the SUSY seesaw models that have been studied for instance in~\cite{Huitu:1994zm,Demir:2008fy,Biondini:2012fv}. The typical signature of the production of $\tilde{\Delta}$ has several leptons and missing transverse energy, analogously to the final states considered in~\cite{ATLAS-CONF-2013-035,CMS-PAS-SUS-12-022}. Similar final state are sensitive to the gauge decays of the tripletino  as for instance $\Delta^{--} \to W^{-} \Delta^{-}$. It is also possible to have resonant pair production of leptons mediated by gauge interactions in the tripletino SUSY decay $\tilde{\Delta}^{++}\to\Delta^{++}\chi^{0}$ for which standard~\cite{Collaboration:2012uq} and dedicated searches~\cite{Babu:2013oz} apply. We expect that with minor modification due to the efficiencies the results of the LHC searches can be applied for this models as well.

The true nature of our model is best displayed when the couplings that dominate the decay of $\Delta$ and $\tilde{Q}$ are either $f_{q}$ or $\lambda$. As in the case of the decay of non-SUSY particles described above, when the triplet $\Delta$ does not decay to leptons it produces heavy quarks (or SUSY partners of them) so it is best to start the discussion for $M_{\Delta}>M_{Q'}$. In this case the decay 
\begin{equation}
\tilde{\Delta}\to q \,\tilde{Q^{'}}
\end{equation}
is dominant mode when $f_{q}\gg f$. In fact the other SUSY mode $\tilde{\Delta}\to \tilde{q} \,Q^{'}$ is likely to be suppressed, or even inaccessible, because of kinematics, as both $\tilde{q}$ and $Q^{\prime}$ are heavy states.
At this point the resulting signature is determined by the decay of $Q^{\prime}$. For $f_{q}\gg\lambda$ the hierarchy of couplings can win over the price of emitting an off-shell $\Delta$ and the resulting decay is 
\begin{equation}
\tilde{Q}^{\prime}\to \tilde{q} \Delta^{*} \to \met + \jets + \leptons\,, \label{susyQtoqDelta}
\end{equation}where the off-shell nature of the $\Delta$ implies non-resonating leptons in the final state. A similar final state is obtained when $Q^{'}$ decays in the SUSY partner final state $\tilde{Q}^{\prime}\to q \tilde{\Delta}^{*}$.

When the coupling $\lambda$  dominates the decay of $\tilde{Q^{'}}$ we have 
\be
\tilde{Q'}\to\tilde{u}^{c} H_{d} \to \met + \jets + H_{d}\,, \label{susyQtoqHd}
\ee
where, depending on the charge of $Q'$, the resulting state in $H_{d}$ decays to $bb$ or $tb$. The decay in the higgsino final state is less well defined because the the SUSY partner $\tilde{H}_{d}$ decays are more model dependent. Therefore we leave them unspecified here, although we expect them to be similar to those included in a inclusive search of for the final states of \eq{susyQtoqHd}.

In principle the decay of $Q^{\prime}$ can result in MSSM squarks and gauginos production via the decay $\tilde{Q}^{\prime}\to \tilde{q} V$ and  $\tilde{Q}^{\prime}\to q \chi$, however these decays are suppressed by the mixing between exotics and SM quarks and therefore they usually have little importance.

The discussion is very similar for the case $M_{\Delta}<M_{Q'}$, that essentially differs from the case  $M_{\Delta}<M_{Q'}$ for the effect of off-shell phase-space factors. This has important implications for the case of the chain \eq{susyQtoqDelta} where the presence of resonating leptons can make a big difference for the experimental searches.

A summary of the decay of SUSY states is presented in Table~\ref{tab:summarydecaysusy}.

\bigskip

A comment on the bounds from standard SUSY searches in our model is in order. In fact we have new sources of SUSY particles and new interactions that in principle can change the bounds for our model. Generically we expect the current SUSY searches to give more stringent bounds in our model on the existence of super-partners. In fact in our model there are extra sources of squarks such as for instance the decays $\tilde{Q}^{\prime}\to\tilde{q}\Delta$  and $\tilde{\Delta} \to \tilde{q} Q^{\prime}$. In most cases these decays have small branching fractions, because the same particles can decay without suppression in other modes that have at least one light final state, hence they systematically beat the modes containing a squark that have always two heavy modes. For this reason we expect only a modest overproduction of quarks and therefore limits are only mildly affected by the new dynamics of our model.

In principle the decay of MSSM states receive corrections from the existence of new states and new interactions that characterize our model. A particularly relevant case is that of the squarks that can decay $\tilde{q}\to \Delta \tilde{Q}^{\prime}$ and  $\tilde{q}\to \tilde{\Delta} Q^{\prime}$, that in general gives rise to more lepton-rich final states than the usual squark decays of the MSSM. From this we expect that a slightly harder bounds from the LHC searches could apply to our model. However, a precise determination of the new bound requires a detailed study. In fact there are other effects that contribute against the isolation of the signal with the strategies adopted in current searches.
For instance our final states typically result in a larger multiplicities and this is not always beneficial for the identification of new physics, as in general the individual objects in high multiplicity events tend to be softer.


\section{New contributions to neutrinoless double beta decay from 5/3 charge quarks\label{sec:doublebeta}} 
\begin{figure}[t]
\begin{center}  
\includegraphics[width=0.5 \linewidth]{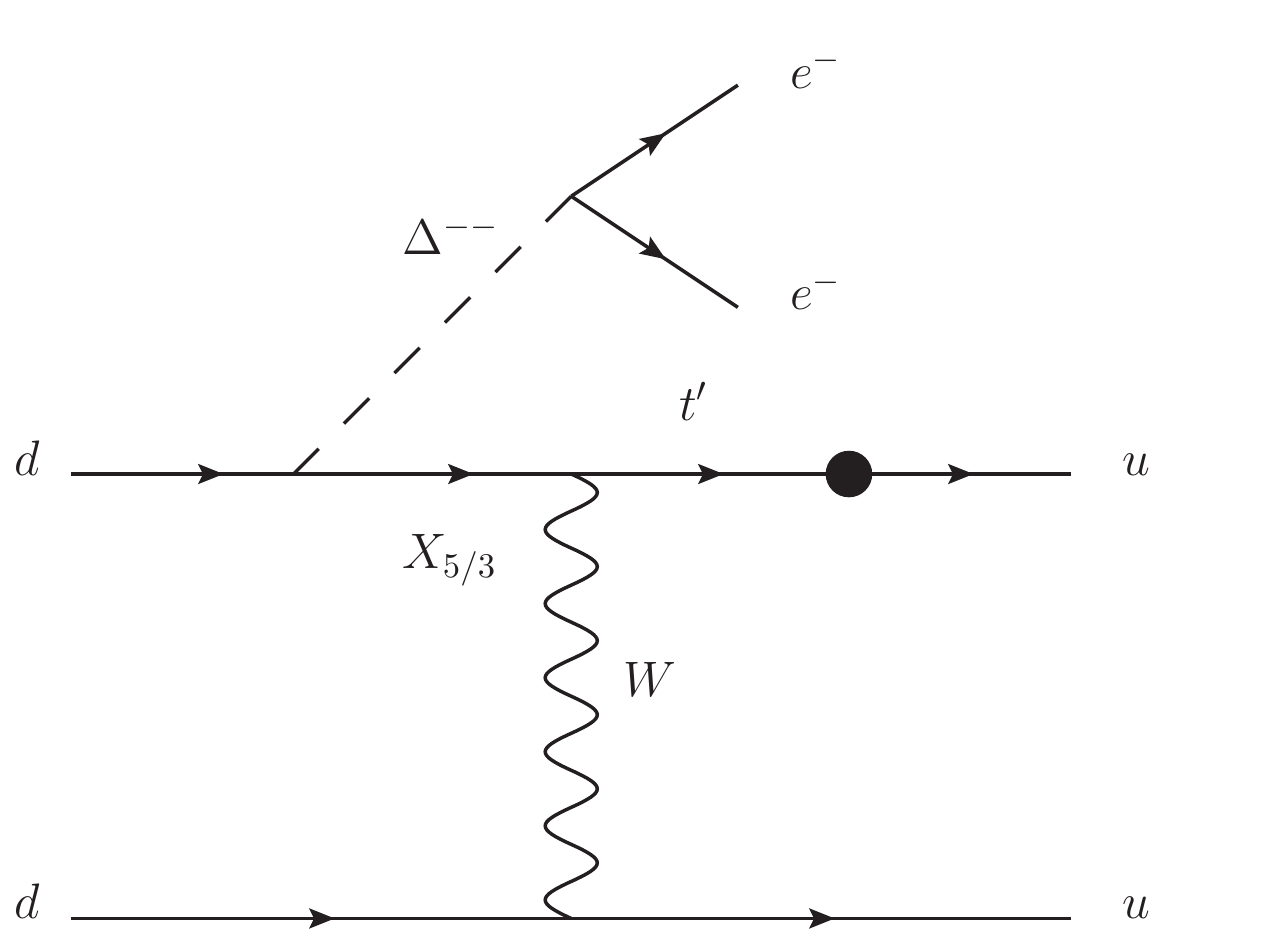}
\end{center}
  \caption{New contribution to neutrinoless double beta decay mediated by $X_{5/3}$ and $\Delta^{--}$.\label{neutrinoless}}
\end{figure}
{Neutrinoless double beta decay~\cite{rodejohann} is a powerful probe of new physics that gives rise to violations of  lepton number. Although the experiments probe relatively low energies in the domain of nuclear physics, the intensity and the precision of these experiments is such as to probe physics at much higher scales. For instance ascribing the $0\nu\beta\beta$ transition to a dimension-9 operator the current generation of experiments is sensitive to operators suppressed by roughly $(1/\TeV)^{5}$. This means that neutrinoless double beta decay can also probe models of EWSB when the  breaking of the gauge symmetry of the SM is in some manner entangled with the breaking of lepton number. Electroweak models for the generation of neutrino masses are therefore an excellent target for $0\nu\beta\beta$ experiments. Our model is no exception to this statement, as we have demonstrated that, not only the neutrino mass is ultimately generated by the VEV of Higgs particles, but also we have found that the overall scale of neutrino masses is set by the scale of SUSY breaking, hence by the electroweak scale.}

In our model there is a new short-range contribution  to neutrino-less double beta decay that involves the couplings $f_q$, $f$, and $\lambda$ via the diagram in Fig.~\ref{neutrinoless}. Remarkably it involves the simultaneous propagation of an exotic quark and a doubly  charged scalar. {This diagram gives rise to $0\nu\beta\beta$ transition mediated by an effective operator of dimension 9, whose strength is partly correlated to the overall scale of neutrino mass. To the best of our knowledge this short-range contributions has not been explicitly spelled out before, although the resulting dimension-9 operator that mediates double beta decay and a possible set of mediators has appeared in studies of the possible decompositions of the $0 \nu\beta \beta$ operator~\cite{Bonnet:2013ys}.}
Therefore this type of contribution appears to be characteristic of seesaw models that have heavy exotic quarks in their spectrum.
{The presence of this new characteristic interplay between particles in the reach of collider and low energy processes definitively constitute further motivation to study our model.}

We estimate that the new contribution to neutrinoless double beta decay mediated by $X_{5/3}$  results in  a matrix element }
\begin{eqnarray}
M^{X_{5/3}}_{\beta\beta}\simeq \frac{G_F}{2} \frac{f_{q_{1}}\lambda_{1} f_{ee}}{M^2_\Delta M_{Q'}} \frac{v_{d}}{M_{Q'}} e^T_LCe_L \bar{u}(1+\gamma_5)\gamma_\alpha d \,\,\bar{u}\gamma^\alpha (1+\gamma_5)d\,,
\end{eqnarray}
where we treat collectively the mass of $t'$ and $X_{5/3}$ by denoting them as  $M_{Q'}$. {We remark that the rate for double beta neutrinoless decay depends on combination of the new couplings $f,\lambda$, and $f_{q}$ that is in general different from the combination $\yeff$ that determines the mass scale of the neutrino. This is remarkable in that it allows to disentangle the rate of the neutrinoless double beta decay from the absolute scale of neutrino masses.} Using eqs.~(\ref{mudeltasum})-(\ref{mneutrino}) the coefficient of this matrix element can be written in terms of the neutrino mass so that
\begin{equation}
M^{X_{5/3}}_{\beta\beta}\simeq G_F\, 8\pi^{2} \,  {\frac{m_{\nu}}{m_{t} M^{3}_{Q'}}}\frac{\lambda_{1}f_{q_{1}}}{\lambda_{3} f_{q_{3}}}\, e^T_LCe_L \bar{u}(1+\gamma_5)\gamma_\alpha d \,\,\bar{u}\gamma^\alpha (1+\gamma_5)d\,,
\end{equation}
{where we assumed the coupling $f$ to not be hierarchical in flavor space, as suggested by the large neutrino mixing, and we also assumed the sum in eq.(\ref{mudeltasum}) to be dominated by the third generation, i.e. $\yeff \simeq \lambda_{3}f_{q_{3}} y_{t}\geq \lambda_{i}f_{q_{i}} y_{u_{i}}$ for $i=1,2$.}
This gives a coefficient that scales like
\begin{equation}
10^{-15} G_{F}^{2} {\frac{\lambda_{1}f_{q_{1}}}{\lambda_{3} f_{q_{3}}}} 
\left(\frac{m_{\nu}}{0.1 \ev}\right)\left( \frac{1 \tev}{M_{Q'}}\right)^{3} \gev^{-1}\,.\label{ratedoublebeta}
\end{equation}
{Note that the smallness of $\mu_{\Delta}$ in eq.(\ref{mneutrino}) allows for $\lambda_3 f_{q3}$ to be small and still to dominate the generation of neutrino masses. Considering only the first and third generation involved here one finds that the third generation dominates in the neutrino mass generation as long as 
\begin{equation}
\frac{\lambda_{3}f_{q_{3}}}{\lambda_{1}f_{q_{1}}} > \frac{y_{u}}{y_{t}} \simeq 10^{-5} \,.
\end{equation}}

Current experimental bounds on the life time for this process roughly translate into bound on the coefficient of this operator to be less than about $G^2_F \times 10^{-8}$ GeV$^{-1}$~\cite{rodejohann,Beringer:1900zz}. 
{Thus there are regions of the parameter space of the model where the double beta amplitude can be near the current upper limit. {For instance, if one assumes safe couplings for precision flavor and electroweak observables, say $\lambda_{1,2}\sim0.1$ and $f_{q_{1,2}}\sim0.1$, taking an ``inverse'' hierarchy for the couplings $\lambda_{i}$ and $f_{q_{i}}$, say $f_{q_{3}}\sim\lambda_3\sim 10^{-4}$ or so, the double beta amplitude for TeV heavy quarks is only about one order of magnitude below the current limit and at the same time eq.(\ref{mneutrino}) predicts sub-eV masses for the neutrinos. Therefore it is possible to test this model through future searches for neutrinoless double beta decay.} {We stress that such hierarchy of the couplings $f_{q_{i}}$ and $\lambda_{i}$ can be cross-checked because it is observable in the lack of heavy flavors in the decay of the heavy quarks summarized in the Tables~\ref{summarydecay} and~\ref{tab:summarydecaysusy}.}

{Furthermore we find that the signal in neutrinoless double beta decay is correlated in a peculiar way to  signals in deep inelastic scattering (DIS) experiments. In fact at momentum scale of order $m_{W}$ the effective operator that generates $0\nu\beta\beta$ in our model opens up leaving a contact interaction  $udWee$ from the interactions of the heavy degrees of freedom $X_{5/3}$ and $\Delta^{++}$. This interaction mediates transitions 
\begin{equation}
e^{-}u\to d \ell^{+}W^{-}\label{dissignal}\,,
\end{equation}
which gives rise to a hard lepton of charge opposite to that of the beam in the DIS experiment. This process is very interesting. In fact it can in principle be used to test the flavor structure of the $f$ coupling responsible for the seesaw. Furthermore, the rate of this reaction (especially for the positron flavor in the final state) is tightly correlated to the rate of $0\nu\beta\beta$. Therefore such DIS experiment allows to test the fraction of the $0\nu\beta\beta$ rate that is due long-range contributions such as the neutrino mass versus the fraction that comes from short-range contributions such as the one of our model in Figure~\ref{neutrinoless}. 
Similar correlations exists in model of leptonic number violating R-parity violating SUSY \cite{Butterworth:1993pb,Aktas:2004ij}, where, however, sleptons are expected in place of the $W$ boson of our process eq.~(\ref{dissignal}). Discerning $W$ bosons from sleptons in DIS final states, of course, offers further information on the actual origin of a possible $0\nu\beta\beta$ signal observed in the next round of experiments.
}

\section{Other implications of the model \label{other}} 

{The presence of new quarks that mix with the SM quarks gives rise to several changes in the phenomenology at energies even much below the mass the of the new quarks.}

{For instance, the couplings $f_{q}$ and $\lambda$ have a flavor index and therefore these couplings can lead  not only to departure from CKM unitarity (which are less than a percent level), but they can also lead to flavor changing neutral currents (FCNC)  e.g. $K-\bar{K}$ oscillations mediated by box graphs involving the exchange of $\Delta$, $H_{d}$ and $Q^{'}$. However, this contribution to the $\Delta S=2$ effective Hamiltonian comes with a strength $\sim \frac{f^{2}_{q_{1}}f^{2}_{q_{2}}}{16\pi^{2}M_{Q}^{2}} \simeq 10^{-6} \TeV^{-2}$ for $f_{q}\sim 0.1$ and $M_{Q}\sim 1\TeV$, which is right at the current bound for chirality and CP conserving transitions~\cite{Isidori:2010zr}. Similar  contributions also arise for $B$ and $D$ mesons mixing and are more easily compatible with the bounds. Along the same lines one can find analogous conclusions for to the case where only the coupling $\lambda$ is used in the box diagram.}
{Since chirality flipping operators have in general much more stringent bounds~\cite{Isidori:2010zr} it is worth remarking that in our model it is likely that this kind of operators are generated with an extra suppression with respect to the chirality conserving ones. In fact these chirality flipping operators can be generated only using simultaneously both the couplings $\lambda_{i}$ and $f_{q_{i}}$. As displayed in eq.~(\ref{mneutrino}) we need the product of two couplings to be $\lambda_{3}f_{q_{3}}\sim 10^{-7}$ to reproduce the overall scale of neutrino masses. Therefore it is natural to assume that at least one of the two couplings must be small, such that the main effect in FCNC comes from box diagrams where only the large coupling  appears. Alternatively, chirality flipping operators arise from mixed box diagrams with one $W$ boson and one new scalar, e.g. $\Delta^{+}W$  box graphs. These are in general  suppressed as they involve small VEVs.}

{Furthermore, new quarks with same electric charge and different $SU(2)$ quantum numbers than the ordinary quarks modify the couplings of the SM quarks to the weak gauge bosons. Especially the couplings of the $Z$ boson are very accurately measured and there are constraints on the amount of mixing that can be tolerated. As discussed in Section~\ref{sec:mixing}, in our model the mixing between SM and new quarks is in general small, due to the smallness of the VEVs responsible for the mixing, that are $v_{d}$ and $v_{\Delta}$ for the $t_{R}$ and $t_{L}$ mixing, respectively. The new quarks also contribute to oblique parameters $S$ and $T$ via their mixing with the SM quarks, which provide yet another constraint. 
Recently these issues have been analyzed in the light of the most up to date results on precision observables and direct searches of heavy quarks~\cite{Aguilar-Saavedra:2013ly} finding that for small mixings as in our model the prediction of the electroweak observables are generically in agreement with the experiment.}

{From the above discussion it is clear that precision flavor and electroweak observable do not deviate significantly from the SM prediction once the new particles have TeV mass and the new couplings $\lambda$ and $f_{q}$ are of the order 0.1 or less. 
Incidentally we remark that, unless the model is complicated such to have cancellations in the BSM contribution to flavor and electroweak precision observables~\cite{Atre:2011wd}, this implies that the single production at colliders of the new states through the couplings $\lambda$, $f_{q}$ is generically much suppressed with respect to the pair production via gauge interactions.}

\section{Conclusions \label{conclusions}} 

We have presented a model of physics of the electroweak scale where the weak scale is originated from the breaking of supersymmetry. At the same time in our model the breaking of supersymmetry triggers the generation of neutrino masses through a mechanism much like the seesaw of type-II. Our model is distinct from the usual SUSY seesaw in that the size of the trilinear interaction of the seesaw triplet and the Higgs doublets, a term $\Delta H_{d}^{*}H_{u}$ in our case, is not put by hand to a tuned value, but rather it is dynamically connected to the scale of the breaking of SUSY, hence to the electroweak scale.
The connection between the scale of generation of neutrino mass and that of stabilization of the Higgs potential is embodied by the diagram of Figure~\ref{oneloop}. To establish this link between the weak scale and neutrino mass generation is necessary to extend the SUSY type-II seesaw model with a new quark of hypercharge 7/6, whose interactions, upon SUSY breaking, make the neutrino mass non-vanishing.

Very remarkably, our mechanism for the generation  of the $\Delta HH$ term of the type-II seesaw model has considerably less fine-tuning that the standard mechanism.  As shown in eq.~(\ref{mneutrino}), in our model the smallness of this trilinear term, and therefore of the neutrino mass, is ascribed to the smallness of three different couplings and not just to the tuning of a single parameter of the Lagrangian. These new couplings are $\lambda$, $f_{q}$, $f$ of the model of eq.~(\ref{susymodel}) and they all represent new Yukawa interaction in the superpotential among MSSM states and states beyond the MSSM. 

It is extremely interesting to notice that to provide the type-II seesaw model with such a natural origin for the $\Delta HH$ term it has been necessary to introduce an exotic hypercharge field. This field has the property that it allows us to write  interactions for the triplet $\Delta$ that extend those of the minimal SUSY type-II seesaw model. These interactions are used to generate radiatively the mass of the neutrinos, and therefore they are the very core of our model.

We have analyzed the observable consequences of the new interaction beyond the generation of neutrino mass. We found two very interesting consequences of this new dynamics. 
We have observed that the dynamics of our new quarks is markedly different from that of minimal models of heavy quarks. In particular we have found significant deviations from the picture of heavy quarks that decay mostly in Goldstone bosons and SM quarks, {\it i.e.} $Q_{H}\to q_{SM} Z,h,W$. The several decay patterns and the corresponding estimates of the decay widths are summarized in Table~\ref{summarydecay} and Table~\ref{tab:summarydecaysusy}. Among all the decays, those mediated by the coupling $\lambda$ stand out as the most different from the usual $Q_{H}\to q_{SM} Z,h,W$ picture. In fact there are scenarios where the decays of the heavy quark can have very modest leptonic activity, as for the decays eqs.(\ref{tprimelambdadecayH}),(\ref{tprimelambdadecayA}),(\ref{lambdadecayX}). The lack of leptons in these signature may significantly impact on the bound on the mass of the existence of heavy quarks, that are currently searched only in the decay modes $Q_{H}\to q_{SM} Z,h,W$. We have seen that in our model there are also scenarios where the final states are lepton-rich, as for instance in the case of eq.~(\ref{Xdeltad}) that can give rise to resonating pairs of leptons of same charge. This signal is so spectacular that, reinterpreting the searches for doubly charged scalars~\cite{Collaboration:2012uq}, one could extend the limits on heavy quarks up to about 1 TeV. Therefore our study gives further motivation to enlarge the set of final states considered for LHC searches for heavy quarks. Furthermore our study calls for of a broader interpretation of current searches for SUSY and other BSM scenarios, which in fact are also probing the dynamics of non-minimal heavy quarks.

Additionally we have observed that the matter content  and interactions of the model generates a dimension 9 effective operator that mediates the neutrinoless double beta decay. This fact {\it per se} is quite expected in models of Majorana neutrinos, as the mass of the neutrino breaks lepton number and therefore can be used as a source for $0\nu\beta\beta$. However, we have found that in our model the rate for $0\nu\beta\beta$ is in general only loosely connected to the generation of the neutrino mass. The degree of disentanglement between the scale of neutrino mass and the rate of double beta decay is represented by the ratio $f_{q_{1}}\lambda_{1}/(f_{q_{3}}\lambda_{3})$ in eq.(\ref{ratedoublebeta}). This shows that we can have scenarios in which the double beta decay is mostly due to the short-range exchange of heavy mediators and only sub-dominantly due to long-range effects from neutrino mass. 
This hierarchy of the contributions to $0\nu\beta\beta$ can be attained especially when the couplings of the new heavy quarks exhibit a sort of inverted hierarchy $f_{q_{1}}\lambda_{1}/(f_{q_{3}}\lambda_{3})\gg1$, i.e. the third generation SM quarks are less strongly coupled to the new quarks than what the first generation does. 
Very excitingly the predicted rate for $0\nu\beta\beta$ can be in the reach of the next round of experiments.
In case a signal will be observed, we have stressed that in principle the LHC {and DIS experiments} can cross-check the possible observation of $0\nu\beta\beta$. At the LHC the most important diagnostic is the flavor composition of the final states of the heavy quarks decay eqs.(\ref{tprimelambdadecayH}),(\ref{tprimelambdadecayA}),(\ref{lambdadecayX}). {At DIS experiments one expects to see a signal in final states with a $W$ boson in association with a hard lepton that has electric charge opposite to that of the beam}.

\section*{Acknowledgments} 
RF thanks Kaustubh Agashe, Mario Kadastik, Martti Raidal, and Luca Vecchi for discussions.
This work is supported by the National Science Foundation Grant No.~PHY-0968854.
The work of RF is also supported by the National Science Foundation Grant No.~PHY-0910467, and by the Maryland Center for Fundamental Physics.
RF  thanks the Galileo Galilei Institute for Theoretical Physics for the hospitality and the INFN for partial support during the completion of this work.

\end{document}